\documentclass[floatfix,aps,superscriptaddress,prd,amsmath,nofootinbib,preprintnumbers,onecolumn]{revtex4}

\usepackage{graphicx}
\usepackage{bm}
\usepackage{float}
\usepackage[
colorlinks=true,        
citecolor=blue,         
linkcolor=blue,         
urlcolor=blue           
]{hyperref}             

\newcommand{\nc}{\newcommand*}
\nc{\Om}{\Omega}
\nc{\ogw}{\Omega_{\mathrm{GW}}}
\nc{\rd}{\mathrm{d}}
\nc{\eg}{\textit{e.g.~}}
\nc{\red}[1]{\textcolor{red}{#1}}
\nc{\lvc}{LIGO/Virgo} 

\def\({\left(}
\def\){\right)}
\def\[{\left[}
\def\]{\right]}
\def\e{\begin{equation}}
\def\q{\end{equation}}
\def\m{\begin{eqnarray}}
\def\n{\end{eqnarray}}

\begin{document}

\title{Probing the scalar-induced gravitational waves with the Five-hundred-meter Aperture Spherical radio Telescope and the Square Kilometer Array}

\author{Jun~Li}
\affiliation{Qingdao Key Laboratory of Novel Optoelectronic Devices and Ultrafast Intelligent Manufacturing, School of Mathematics and Physics, Qingdao University of Science and Technology, Qingdao 266061, China.}
\affiliation{CAS Key Laboratory of Theoretical Physics, Institute of Theoretical Physics, Chinese Academy of Sciences, Beijing 100190, China}

\thanks{This work is accepted for publication in \emph{Chinese Physics C}.}

\author{Guanghai Guo}
\affiliation{Qingdao Key Laboratory of Novel Optoelectronic Devices and Ultrafast Intelligent Manufacturing, School of Mathematics and Physics, Qingdao University of Science and Technology, Qingdao 266061, China.}

\author{Pengfei Yan}
\affiliation{Qingdao Key Laboratory of Novel Optoelectronic Devices and Ultrafast Intelligent Manufacturing, School of Mathematics and Physics, Qingdao University of Science and Technology, Qingdao 266061, China.}

\date{\today}

\begin{abstract}

Gravitational wave astronomy presents a promising opportunity to directly observe scalar-induced gravitational waves originating from the early universe. Experiments, including ground-based interferometers like LIGO and Virgo, and the Pulsar Timing Array, such as FAST and SKA, are poised to significantly enhance sensitivity to these gravitational waves. In this paper, we combined Cosmic Microwave Background and Baryon Acoustic Oscillation datasets with upper or lower limits of the stochastic gravitational wave background provided by FAST or SKA, to constrain scalar-induced gravitational waves. To provide a comprehensive forecast, we consider two scenarios at a frequency: one where FAST or SKA does not detect scalar-induced gravitational waves, thereby setting an upper limit on the fractional energy density; and another where these waves are detected successfully, thus establishing a lower limit. In the $\Lambda$CDM+$r$ model, the scalar spectral index of the power-law power spectrum is constrained to $n_s=0.9598^{+0.0013}_{-0.0009}$ from the combinations of CMB+BAO+SKA datasets in the upper limit scenario where scalar-induced gravitational waves propagate at the speed of light. The constraint shifts to $n_s =  0.9697\pm{0.0033}$ in the lower limit scenario. Comparing with the constraint from the combinations of CMB+BAO datasets, the scalar spectral index $n_s$ in the upper limit scenario exhibits significant changes, which could serve as an indicator for detecting scalar-induced gravitational waves. In the $\Lambda$CDM+$\alpha_s$+$r$ model and $\Lambda$CDM+$\alpha_s$+$\beta_s$+$r$ model, the running of the scalar spectral index $\alpha_s$ and the running of the running $\beta_s$ also show notable variations, suggesting potential indicators. The numerical findings clearly demonstrate the impact of the upper and lower limits provided by FAST or SKA.

\end{abstract}

\maketitle

\section{introduction}
Since the first direct detection of gravitational waves by the Laser Interferometer Gravitational-Wave Observatory (LIGO) in 2015 \cite{LIGOScientific:2016aoc}, the field of gravitational-wave astronomy has undergone rapid transformation. This progress is marked by numerous detections, most notably the multi-messenger observation of gravitational and electromagnetic signals from a binary neutron star merger in 2017 \cite{LIGOScientific:2017vwq}. The observational landscape is poised for further expansion with next-generation facilities. Space-based observatories like the Laser Interferometer Space Antenna (LISA) \cite{Caprini:2015zlo} will access new frequency bands and unprecedented sensitivities, enabling the study of previously inaccessible astrophysical and cosmological sources. Concurrently, the nanohertz frequency band is being probed by pulsar timing arrays (PTAs). Major international collaborations, including the Chinese PTA (CPTA) \cite{Xu:2023wog}, the European PTA (EPTA) in conjunction with the Indian PTA (InPTA) \cite{EPTA:2023sfo,EPTA:2023fyk}, the Parkes PTA (PPTA) \cite{Zic:2023gta,Reardon:2023gzh}, and the North American Nanohertz Observatory for Gravitational Waves (NANOGrav) \cite{NANOGrav:2023gor,NANOGrav:2023hde}, are collectively advancing the detection of low-frequency gravitational waves through precision timing of millisecond pulsars. The EPTA second data release incorporates data from 25 pulsars, with a total observational time span of 24.7 years. The PPTA third data release consists of observations of 32 pulsars spanning up to 18 years, while the NANOGrav 15-year dataset includes timing data from 68 pulsars collected over a 16.03‑year baseline. Radio telescopes such as the Five-hundred-meter Aperture Spherical radio Telescope (FAST) \cite{Nan:2011um} and the future Square Kilometre Array (SKA) \cite{Kuroda:2015owv} play crucial supporting roles in this global endeavor, both by discovering and monitoring the pulsars essential for PTAs and by contributing to multi-messenger follow-up observations. The sensitivity curves for FAST and SKA are adopted from \cite{Kuroda:2015owv}, assuming observational spans of 50 and 100 years from 50 and 100 pulsars, respectively. In both cases, the accessible frequency range extends down to the sub‑nanohertz regime, where the lowest observable frequency is set by the inverse of the total observation time.

The detection of gravitational waves has opened a new observational window in astrophysics and cosmology, providing a powerful tool for probing the early Universe and a wide range of astrophysical phenomena. A key focus of this research involves the study of gravitational waves of primordial origin and those induced by scalar perturbations. Theoretical predictions for primordial gravitational waves in modified gravity models are reviewed, with particular emphasis on phenomenological consequences such as modifications to the graviton mass \cite{Brax:2017pzt,Dubovsky:2009xk,Li:2017cds} and the propagation speed of gravitational waves \cite{Cai:2020ovp,Cai:2016ldn,Lin:2016gve,Giare:2020vss,Ezquiaga:2021ler,Li:2024cmk}. Observational searches for these signatures are conducted via measurements of the cosmic microwave background (CMB), pulsar timing arrays, and laser interferometers \cite{Campeti:2020xwn,Li:2019vlb,Li:2021scb}. The connection between such models and the formation mechanisms of primordial black holes is also explored \cite{Li:2018iwg}. Primordial tensor modes originate directly from the quantum fluctuations of the gravitational field during inflation. In contrast, scalar-induced gravitational waves are second-order tensor perturbations, generated by the non-linear coupling of scalar curvature perturbations. A distinct class of stochastic gravitational-wave backgrounds can emerge from such second-order cosmological perturbations, in which scalar modes act as a source for tensor fluctuations, thereby producing a detectable induced gravitational-wave signal \cite{Ananda:2006af,Baumann:2007zm}. In the radiation-dominated era, the spectral properties of these induced waves have been calculated for various primordial power spectra \cite{Kohri:2018awv,Lu:2019sti}. Their observable spectra in different inflationary scenarios \cite{Alabidi:2012ex,Alabidi:2013lya,Di:2017ndc,Xu:2019bdp,Zhou:2020kkf,Osano:2006ew}, and the modulating effects of primordial non-Gaussianity \cite{Unal:2018yaa,Cai:2018dig}, are actively discussed. While detection is challenging due to their inherently second-order and thus suppressed amplitude, typically scaling as the square of the curvature perturbations, they can become comparable to or even dominate over the primordial signal if the primordial perturbations are sufficiently enhanced on small scales \cite{Inomata:2018epa,Inomata:2019ivs,Inomata:2019zqy,Yuan:2019wwo,Cai:2019amo,Matarrese:1997ay,Noh:2004bc,Li:2022avp}. The detection of scalar-induced gravitational waves holds significant promise. These waves offer a direct observational probe of the final stages of inflation and the state of the very early Universe prior to Big Bang Nucleosynthesis. This potential is underscored by studies that derive constraints on the amplitude of primordial density perturbations from induced gravitational wave signals \cite{Assadullahi:2009jc,Assadullahi:2009nf,Li:2021uvn}, and by proposals to measure the propagation speed of these waves through observations \cite{Cai:2019jah,Li:2023uhu,Chen:2024fir}. The propagation speed of gravitational waves constitutes a fundamental issue in gravitational theory. According to general relativity, gravitational waves travel at the speed of light. However, alternative theories of gravity propose modifications to general relativity, and deviations from the luminal speed on cosmological scales could signal modified gravity or new underlying physics. Although the propagation speed of primordial gravitational waves has been thoroughly investigated, the corresponding study for scalar‑induced gravitational waves remains comparatively sparse and calls for more focused research. 

Both primordial gravitational waves and scalar-induced gravitational waves contribute to the generation of a stochastic gravitational wave background spanning a broad spectrum of frequency bands. Importantly, all these gravitational wave observations are sensitive to stochastic gravitational wave background. The stochastic gravitational wave background signal cannot be confidently inferred from current data. Here, we assume the signal could be explained by a stochastic gravitational wave background. To achieve better constraints on primordial gravitational waves or scalar-induced gravitational waves, it is essential to combine observational datasets spanning different frequency bands. The LIGO and Virgo detectors, for instance, cover the high-frequency range from 20 Hz to 1726 Hz. LISA operates in the frequency range from $10^{-4}$ Hz to $1$ Hz, while PTAs detect signals in the low-frequency range from $1.58\times10^{-9}$ Hz to $8.27\times10^{-7}$ Hz. Additionally, FAST spans frequencies from $6.34\times10^{-10}$ Hz to $8.27\times10^{-7}$ Hz, and SKA covers from $3.17\times10^{-10}$ Hz to $8.27\times10^{-7}$ Hz. Combining data from these different instruments across their respective frequency ranges holds the promise of providing deeper insights into the origins and nature of gravitational waves in our universe. In our previous study \cite{Li:2019vlb}, we combined CMB B-mode data from the BICEP2 and Keck array through 2015 reason \cite{BICEP2:2018kqh}, along with the null search results of the stochastic gravitational wave background from LIGO and Virgo detectors, to establish the constraints on primordial gravitational waves. Additionally, we projected potential improvements using future gravitational wave experiments such as LISA, PTA, and FAST, by integrating their data with CMB B-mode polarization data.

In this context, we focus on scalar‑induced gravitational waves and assume that the observed stochastic gravitationalwave signal originates from this mechanism. According to observations by the Planck satellite \cite{Planck:2018vyg} and Baryon Acoustic Oscillation (BAO) measurements \cite{Beutler:2011hx, Ross:2014qpa, BOSS:2016wmc}, we assume that the curvature power spectrum can be extrapolated from CMB to PTA scales without significant running. The fractional energy density of scalar-induced gravitational waves around a frequency of $10^{-10}$ Hz is estimated to be approximately $10^{-17}$. When comparing this estimate with the sensitivity curves for frequency and fractional energy density of detectors such as LIGO, Virgo, and LISA, it becomes evident that these instruments cannot effectively probe such minuscule scalar-induced gravitational waves. Current PTA observations are also unable to effectively probe such faint scalar-induced gravitational waves in the nanohertz regime. However, the sensitivity curves of the radio telescopes FAST and SKA extend into the sub‑nanohertz band, making them promising instruments for constraining scalar‑induced gravitational waves. This study adopts a sensitivity‑driven threshold analysis, in which projected limits on the gravitational‑wave energy density from the FAST and SKA sensitivity curves are used as constraints, rather than performing a full Bayesian parameter estimation with a pulsar‑timing‑array likelihood function. We adopt the sensitivity curves for FAST and SKA from \cite{Kuroda:2015owv}, assuming observational time baselines of 50 and 100 years, respectively. These long baselines are considerably extended compared to those in more recent FAST/SKA PTA forecasts studies, which assume shorter observational spans and may therefore remain insufficient for detecting the faint scalar‑induced gravitational‑wave signal.
To derive constraints on scalar-induced gravitational waves, we combined CMB data from the Planck satellite \cite{Planck:2018vyg}, the BICEP and Keck Array through the 2018 Observing Season (BK18) \cite{BICEP:2021xfz}, and Baryon Acoustic Oscillation (BAO) measurements \cite{Beutler:2011hx, Ross:2014qpa, BOSS:2016wmc} including the latest DESI Data Release 2 results \cite{DESI:2025zgx}, along with upper or lower limits of the stochastic gravitational wave background provided by FAST or SKA, to establish the constraints on scalar-induced gravitational waves.

\section{the scalar-induced gravitational waves}
In the conformal Newtonian gauge, the metric perturbation about the Friedmann-Robertson-Walker (FRW) background is typically expressed as
\e
\mathrm{d}s^2=a^2\left\{-(1+2\Phi)\mathrm{d}\eta^2+\left[(1-2\Phi)\delta_{ij}+\frac{h_{ij}}{2}\right]\mathrm{d}x^i\mathrm{d}x^j \right\},      \label{metric}
\q
where $\eta$ denotes conformal time, $a(\eta)$ represents the scale factor of the FRW universe, $\Phi$ is the scalar perturbation representing the gravitational potential and $h_{ij}$ represents the tensor perturbation, which is transverse and traceless. We neglect the vector perturbation, the first-order gravitational waves and the anisotropic stress. In the Fourier space, the tensor perturbation $h_{ij}$ is expressed as
\e
h_{ij}(\eta,\mathbf{x})=\int\frac{\mathrm{d}^3k}{(2\pi)^{3/2}}\Bigg(e_{ij}^{+}(\mathbf{k})h_{\mathbf{k}}^+(\eta)+e_{ij}^{\times}(\mathbf{k})h_{\mathbf{k}}^{\times}(\eta)\Bigg)e^{i\mathbf{k}\cdot\mathbf{x}},
\q
where the plus and cross polarization tensors can be expressed as
\e
e_{ij}^{+}(\mathbf{k})=\frac{1}{\sqrt{2}}\Bigg(e_i(\mathbf{k})e_j(\mathbf{k})-\bar{e}_i(\mathbf{k})\bar{e}_j(\mathbf{k})\Bigg),\quad
e_{ij}^{\times}(\mathbf{k})=\frac{1}{\sqrt{2}}\Bigg(e_i(\mathbf{k})\bar{e}_j(\mathbf{k})+\bar{e}_i(\mathbf{k})e_j(\mathbf{k})\Bigg),
\q
the normalized vectors $e_i(\mathbf{k})$ and $\bar{e}_i(\mathbf{k})$ are mutually orthogonal and orthogonal to $\mathbf{k}$. The tensor equation of motion for $h_{ij}$ can be straightforwardly derived from the perturbed Einstein equations up to second order. Scalar perturbations couple to tensor perturbations in the second-order equation. For this analysis, we also consider the propagation speed of scalar-induced gravitational waves within the framework of a FRW universe, and assume the speed of scalar-induced gravitational waves to be a constant parameter. The equation governing induced gravitational waves, with $\Phi_{\mathbf{k}}$ as the source, is given by
\e
h_{\mathbf{k}}^{\prime\prime}(\eta)+2\mathcal{H}h_{\mathbf{k}}^\prime(\eta)+c_g^2k^2h_{\mathbf{k}}(\eta)=4S_{\mathbf{k}}(\eta), \label{gw}
\q
where the prime denotes derivative with respect to conformal time, $\mathcal{H}=a^\prime/a=aH$ represents the conformal Hubble parameter, and $c_g$ denotes the speed of scalar-induced gravitational waves. In this paper, $c_g$ appears to be treated as a phenomenological parameterization of
the propagation speed, rather than being embedded in a fully specified theoretical framework. The source term is given by
\e
S_{\mathbf{k}}=\int\frac{\mathrm{d}^3q}{(2\pi)^{3/2}}e_{ij}(\mathbf{k})q_iq_j\Bigg(2\Phi_{\mathbf{q}}\Phi_{\mathbf{k-q}}+\frac{4}{3(1+\omega)}\left(\mathcal{H}^{-1}\Phi^{\prime}_{\mathbf{q}}+\Phi_{\mathbf{q}}\right)\left(\mathcal{H}^{-1}\Phi^{\prime}_{\mathbf{k-q}}+\Phi_{\mathbf{k-q}}\right)\Bigg).
\q
We consider the Green's function method as
\e
h_{\mathbf{k}}(\eta)=\frac{4}{a(\eta)}\int^{\eta}{\mathrm{d}}\bar{\eta}G_{\mathbf{k}}(\eta,\bar{\eta})a(\bar{\eta})S_{\mathbf{k}}(\bar{\eta}),
\q
where $G_{\mathbf{k}}(\eta,\bar{\eta})$ satisfies the equation
\e
G_{\mathbf{k}}^{\prime\prime}(\eta,\bar{\eta})+\left(c_g^2k^2-\frac{a^{\prime\prime}(\eta)}{a(\eta)}\right)G_{\mathbf{k}}(\eta,\bar{\eta})=\delta(\eta-\bar{\eta}).
\q
In the Radiation dominated Universe, the solution of the Green's function is
\e
G_{\mathbf{k}}(\eta,\bar{\eta})=\frac{1}{c_gk}\sin[c_gk(\eta-\bar{\eta})]\Theta(\eta-\bar{\eta}),
\q
where $\Theta$ being the Heaviside step function.

The power spectrum of scalar-induced gravitational waves is defined as
\e
\langle h_{\mathbf{k}}(\eta) h_{\mathbf{k}^{\prime}}(\eta)\rangle=\frac{2\pi^2}{k^3}\delta^{(3)}(\mathbf{k}+\mathbf{k}^{\prime})\mathcal{P}_h(\eta, k),
\q
and the fractional energy density is defined as \cite{Allen:1997ad,Maggiore:1999vm}
\e
\Omega_{\mathrm{GW}}(\eta, k)=\frac{1}{\rho_c}\frac{d\rho_{\mathrm{GW}}}{d\ln k}=\frac{1}{24}\Bigg(\frac{c_gk}{aH}\Bigg)^2\overline{\mathcal{P}_h(\eta, k)},
\q
where $\rho_{\mathrm{GW}}$ is the energy density of the stochastic background of gravitational waves, and $\rho_c$ is the critical energy density.
After calculation, the power spectrum of scalar-induced gravitational waves takes the form
\e
\mathcal{P}_h(\eta, k)=4\int_0^\infty\mathrm{d}v\int_{\vert1-v\vert}^{1+v}\mathrm{d}u\left(\frac{4v^2-(1+v^2-u^2)^2}{4vu}\right)^2I^2(v, u, x)\mathcal{P}_{\zeta}(kv)\mathcal{P}_{\zeta}(ku), \label{ph}
\q
where $\mathcal{P}_{\zeta}(k)$ is the power spectrum of the primordial curvature perturbations, $x\equiv k\eta$, $u=|\mathbf{k}-\tilde{\mathbf{k}}|/k$ and $v=\tilde k/k$. The function $I(v, u, x)$ is defined as
\e
I(v, u, x)=\int_0^x\mathrm{d}\bar{x}\frac{a(\bar{\eta})}{a(\eta)}kG_{\mathbf{k}}(\eta,\bar{\eta})f(v, u, \bar{x}),
\q
where $\bar{x}\equiv k\bar{\eta}$, and $f(v, u, \bar{x})$ comes from the source term
\begin{align}
f(v, u, \bar{x})=&\frac{6(\omega+1)}{3\omega+5}\Phi(v\bar x)\Phi(u\bar x)+\frac{6(1+3\omega)(\omega+1)}{(3\omega+5)^2}\Big(\bar x \partial_{\bar{\eta}}\Phi(v\bar x)\Phi(u\bar x)+\bar x \partial_{\bar{\eta}}\Phi(u\bar x)\Phi(v\bar x)\Big) \nonumber\\
&+\frac{3(1+3\omega)^2(1+\omega)}{(3\omega+5)^2}{\bar x}^2\partial_{\bar{\eta}}\Phi(v\bar x)\partial_{\bar{\eta}}\Phi(u\bar x).
\end{align}
In the radiation-dominated Universe, the functions $f(v, u, \bar{x})$ and $I(v, u, x)$ can be written as our previous study \cite{Li:2023uhu}.

For a power-law scalar power spectrum
\m
\mathcal{P}_{\zeta}(k)&=&A_s\(\frac{k}{k_*}\)^{n_s-1+\frac{1}{2}\alpha_s\ln(k/k_*)+\frac{1}{6}\beta_s(\ln(k/k_*))^2}, \label{eqs:spectrumscalar}
\n
the power spectrum of scalar-induced gravitational waves is given by \cite{Li:2023uhu}
\e
\mathcal{P}_h(\eta, k)=\frac{24Q(c_g, n_s, \alpha_s, \beta_s, k)}{(k\eta)^2}A^2_s\(\frac{k}{k_*}\)^{2\[{n_s-1+\frac{1}{2}\alpha_s\ln(k/k_*)+\frac{1}{6}\beta_s(\ln(k/k_*))^2}\]},
\q
where $A_s$ represents the scalar amplitude of the power-law power spectrum at the pivot scale $k_*=0.05$ Mpc$^{-1}$, $n_s$ denotes the scalar spectral index, $\alpha_s\equiv\mathrm{d} n_s/\mathrm{d}\ln k$ represents the running of the scalar spectral index, $\beta_s\equiv{\mathrm{d}^2n_s}/{\mathrm{d}\ln k^2}$ indicates the running of the running of the scalar spectral index, and $Q(c_g, n_s, \alpha_s, \beta_s, k)$ are the overall coefficients 
\begin{align}
Q(c_g, n_s, \alpha_s, \beta_s, k)=&\frac{1}{12}\int_0^\infty\mathrm{d}v\int_{\vert1-v\vert}^{1+v}\mathrm{d}u \Bigg(\frac{4v^2-(1+v^2-u^2)^2}{4vu}\Bigg)^2 \Bigg(\frac{3(u^2+v^2-3c_g^2)}{4u^3v^3}\Bigg)^2 \nonumber\\
&\Bigg(\Big(-4uv+(u^2+v^2-3c_g^2)\log\left|\frac{3c_g^2-(u+v)^2}{3c_g^2-(u-v)^2}\right|\Big)^2+\pi^2(u^2+v^2-3c_g^2)^2\Theta(u+v-\sqrt{3}c_g)\Bigg) \nonumber\\
&\Bigg(\frac{k}{k_*}\Bigg)^{\frac{1}{2}\alpha_s\ln(uv)+\frac{1}{6}\beta_s\Big((\ln{v})^2+2\ln{v}\ln(k/k_*)+(\ln{u})^2+2\ln{u}\ln(k/k_*)\Big)}\Bigg(uv\Bigg)^{n_s-1+\frac{1}{2}\alpha_s\ln(k/k_*)+\frac{1}{6}\beta_s(\ln(k/k_*))^2}\nonumber\\
&v^{\frac{1}{2}\alpha_s\ln{v}+\frac{1}{6}\beta_s\Big((\ln{v})^2+2\ln{v}\ln(k/k_*)\Big)}u^{\frac{1}{2}\alpha_s\ln{u}+\frac{1}{6}\beta_s\Big((\ln{u})^2+2\ln{u}\ln(k/k_*)\Big)},
\end{align}
which is the integral of $u$ and $v$, and depend on $c_g$, $n_s$, $\alpha_s$, $\beta_s$ and $k$. The numerical values for representative parameter sets are listed in Table 4 of \cite{Li:2023uhu}. The connection between the parameters of power-law spectrum and scalar-induced gravitational waves is established. The fractional energy density becomes
\e
\Omega_{\mathrm{GW}}(\eta, k)=Q(c_g, n_s, \alpha_s, \beta_s, k)A^2_s\(\frac{k}{k_*}\)^{2\[{n_s-1+\frac{1}{2}\alpha_s\ln(k/k_*)+\frac{1}{6}\beta_s(\ln(k/k_*))^2}\]}.  \label{power law2}
\q
Here, we investigate the scalar-induced gravitational waves propagating at the speed of light first. According to Planck+BAO observations \cite{Planck:2018vyg}, the central values are $n_s=0.9647\pm0.0043$, $\alpha_s=0.009\pm0.012$ and $\beta_s=0.0011\pm0.0099$. Although Planck+BAO detectors do not directly probe frequencies near nHz, we assume that the curvature power spectrum can be extrapolated from CMB to PTA scales without significant running. The fractional energy density of the scalar-induced gravitational waves at the frequency of $10^{-10}$ Hz would be of the order $10^{-17}$.

\begin{figure}[htb]
\centering
\includegraphics[width=10cm]{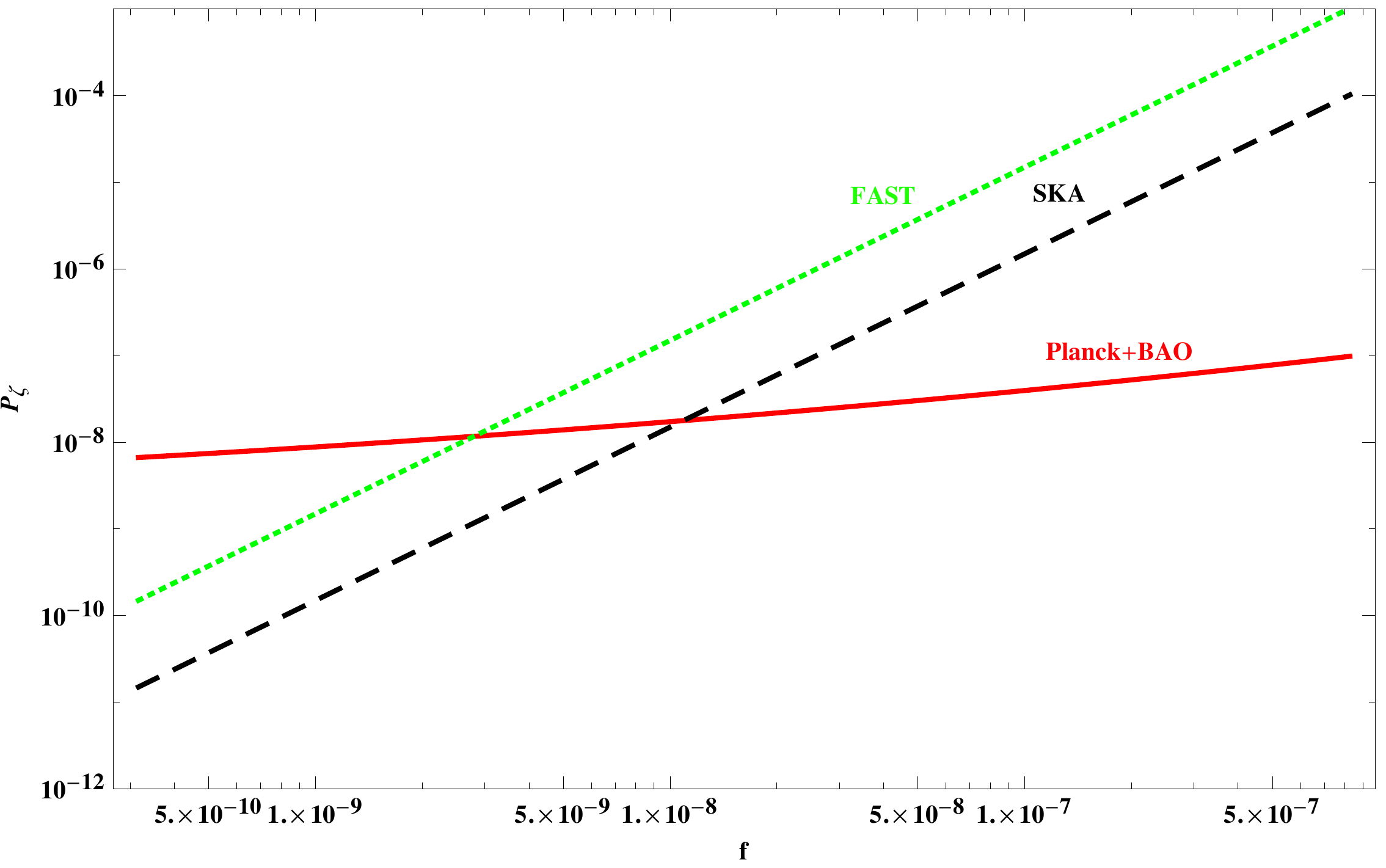}
\caption{The power-law scalar power spectrum $\mathcal{P}_{\zeta}$ from Planck+BAO, FAST and SKA detectors across all scales.}
\label{figure5}
\end{figure}

The gravitational-wave detections also show the sensitivity curves of frequency and $\Omega_{\mathrm{GW}}$ in the detectible ranges which can be used to find scalar-induced gravitational waves. To comprehensively forecast, we consider two scenarios at a frequency: one where gravitational-wave detections cannot detect scalar-induced gravitational waves, constraining the fractional energy density by an upper limit. Alternatively, we explore a scenario where gravitational-wave detections can detect scalar-induced gravitational waves, thereby constraining the fractional energy density by a lower limit. Combining these limits with Eq.~(\ref{power law2}), we can obtain the constraints on the power spectrum of the primordial curvature perturbations. Comparing with LIGO, Virgo, LISA, and PTA detectors, the sensitivity curve of FAST and the energy density fraction $\Omega_{\mathrm{GW}}$ of the scalar-induced gravitational waves would intersect at the frequency of $3\times10^{-9}$ Hz which exhibits in Figure 3 of \cite{Li:2022avp}. It illustrates that FAST detector reaches the region predicted by Planck+BAO observations and divides it into two parts at frequencies below $3\times10^{-9}$ Hz which is also obvious in Figure \ref{figure5}. We plot the power-law scalar power spectrum $\mathcal{P}_{\zeta}$ from Planck+BAO, FAST and SKA detectors across all scales relevant for this work and illustrate where the additional constraints from FAST and SKA arise. The region above the dashed green curve represent the lower limit region for FAST, and the region under the dashed green curve represent the upper limit region. According to the sensitivity curve of FAST \cite{Kuroda:2015owv}, 
\begin{align}
\Omega_{\mathrm{GW}}&=\frac{2\pi^2}{3H_0^2}f^2h_c^2(f),\quad h_c(f)=1.5\times10^{-17}\times\frac{f}{3.17\times10^{-8}},
\end{align}
the fractional energy density of the scalar-induced gravitational waves at the frequency of $6.34\times10^{-10}$ Hz would be of the order $10^{-19}$. Correspondingly, the upper limit is $\Omega_{\mathrm{GW}}<10^{-19}$, and the lower limit is $\Omega_{\mathrm{GW}}>10^{-19}$ at the frequency of $6.34\times10^{-10}$ Hz which are used in the following section. Similarly, for SKA, 
\begin{align}
\Omega_{\mathrm{GW}}&=\frac{2\pi^2}{3H_0^2}f^2h_c^2(f),\quad h_c(f)=1.5\times10^{-18}\times\frac{f}{3.17\times10^{-8}},
\end{align}
the upper limit is $\Omega_{\mathrm{GW}}<10^{-22}$, and the lower limit is $\Omega_{\mathrm{GW}}>10^{-22}$ at the frequency of $3.17\times10^{-10}$ Hz.
The values of $\Omega_{\mathrm{GW}}$ adopted for FAST and SKA are obtained from the corresponding sensitivity curves. Through Eq.~(\ref{power law2}), the fractional energy density $\Omega_{\mathrm{GW}}$ is connected to the parameters of power-law spectrum. Since the spectral parameters $\{A_s, n_s, \alpha_s,\beta_s\}$ influence the fractional energy density, they are sensitive to the upper or lower limit of FAST. So we could expect that the sensitivity curve of FAST leads to distinct constraints on the parameters of power-law spectrum. For FAST and SKA, the accessible frequency range extends down to the sub‑nanohertz regime, the order of fractional energy density change obviously. It is therefore anticipated that measurements in this regime will yield characteristically distinct limits on the spectral index parameters.

Recently, the PTA community has achieved significant progress, with multiple collaborations presenting compelling evidence supporting the existence of a stochastic signal in the frequency range of approximately 1-100 nHz. The presence of the currently observed foreground will greatly limit the constraining power of future observations on subdominant stochastic gravitational wave background \cite{Babak:2024yhu}. For FAST and SKA, the observation span is assumed as 50 years and 100 years respectively \cite{Kuroda:2015owv}. The accessible frequency range extends below nHz. In this work, we consider the upper limit or lower limit at frequencies below nHz which still have no observation. Currently, we can neglect the impact of possible foregrounds at frequencies below nHz. It will be necessary to consider the foreground noise in the future when we obtain the original observation at frequencies below nHz and search for subdominant stochastic gravitational wave background. In the future, when PTA observations are able to detect stochastic signals in the sub‑nHz band, a measured energy density significantly higher than $10^{-17}$ could be attributed to non-SIGW origins as is currently done for existing signals which aligns with the predictions of this work. If the detected energy density is comparable to the predicted SIGW level, the signal might originate from scalar‑induced gravitational waves, a hypothesis that could be tested using the scalar spectral index as a distinguishing indicator. Given that multiple source populations likely contribute to the stochastic background, removing the brighter, astrophysical foreground components may reveal a residual compatible with SIGWs. As detector sensitivities improve, a clear SIGW signature is expected to emerge in future data, underscoring the importance of monitoring $n_s$ as a key diagnostic in separating cosmological from astrophysical gravitational‑wave backgrounds.

\section{the constraints on scalar-induced gravitational waves from the combinations of CMB+BAO with upper or lower limits provided by FAST or SKA}
\begin{figure}[htb]
\centering
\includegraphics[width=17.5cm]{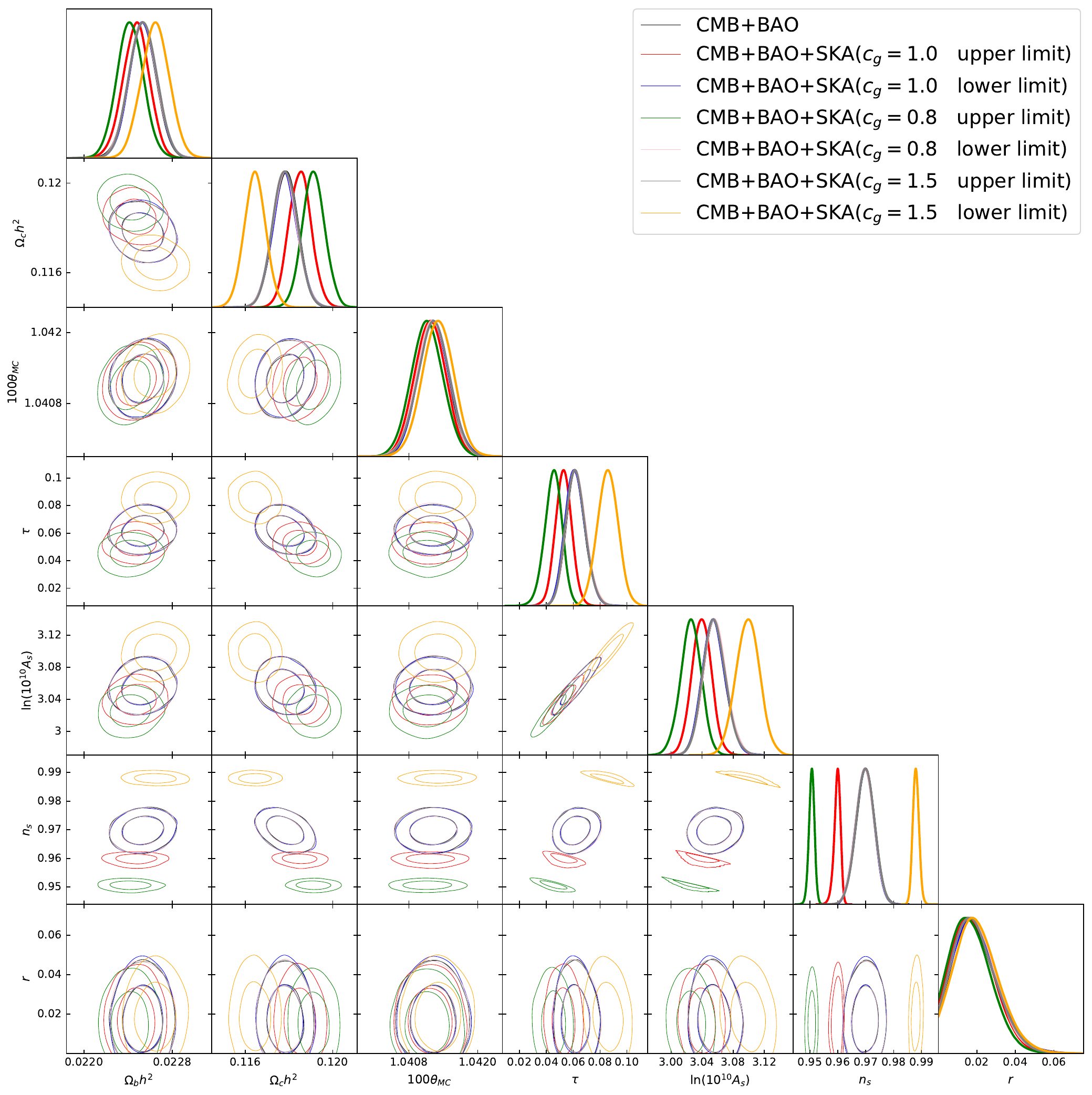}
\caption{The contour plots and likelihood distributions of cosmological parameters in the $\Lambda$CDM+$r$ model are shown at the $68\%$ and $95\%$ confidence levels, derived from the combinations of CMB+BAO, CMB+BAO+SKA($c_g=1.0\,$upper limit), CMB+BAO+SKA($c_g=1.0\,$lower limit), CMB+BAO+SKA($c_g=0.8\,$upper limit), CMB+BAO+SKA($c_g=0.8\,$lower limit), CMB+BAO+SKA($c_g=1.5\,$upper limit), and CMB+BAO+SKA($c_g=1.5\,$lower limit) datasets, respectively. The filled lines in the likelihood distributions represent the constraints from CMB+BAO+SKA datasets. The dashed lines in the likelihood distributions represent the constraints from CMB+BAO datasets.}
\label{figure1}
\end{figure}

\begin{figure}[htb]
\centering
\includegraphics[width=17.5cm]{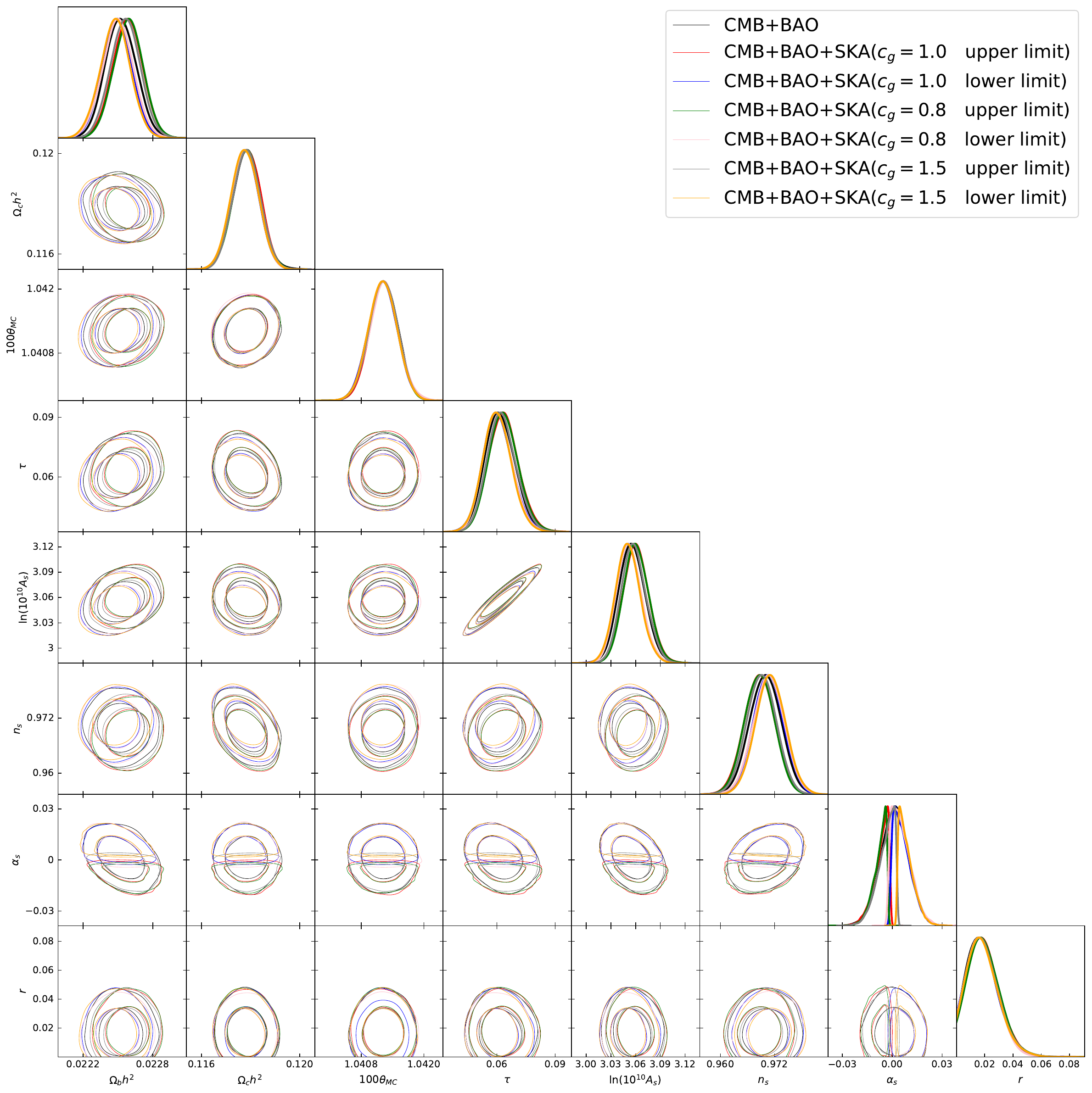}
\caption{The contour plots and likelihood distributions of cosmological parameters in the $\Lambda$CDM+$\alpha_s$+$r$ model are shown at the $68\%$ and $95\%$ confidence levels, derived from the combinations of CMB+BAO, CMB+BAO+SKA($c_g=1.0\,$upper limit), CMB+BAO+SKA($c_g=1.0\,$lower limit), CMB+BAO+SKA($c_g=0.8\,$upper limit), CMB+BAO+SKA($c_g=0.8\,$lower limit), CMB+BAO+SKA($c_g=1.5\,$upper limit), and CMB+BAO+SKA($c_g=1.5\,$lower limit) datasets, respectively. The filled lines in the likelihood distributions represent the constraints from CMB+BAO+SKA datasets. The dashed lines in the likelihood distributions represent the constraints from CMB+BAO datasets.}
\label{figure2}
\end{figure}

\begin{figure}[htb]
\centering
\includegraphics[width=17.5cm]{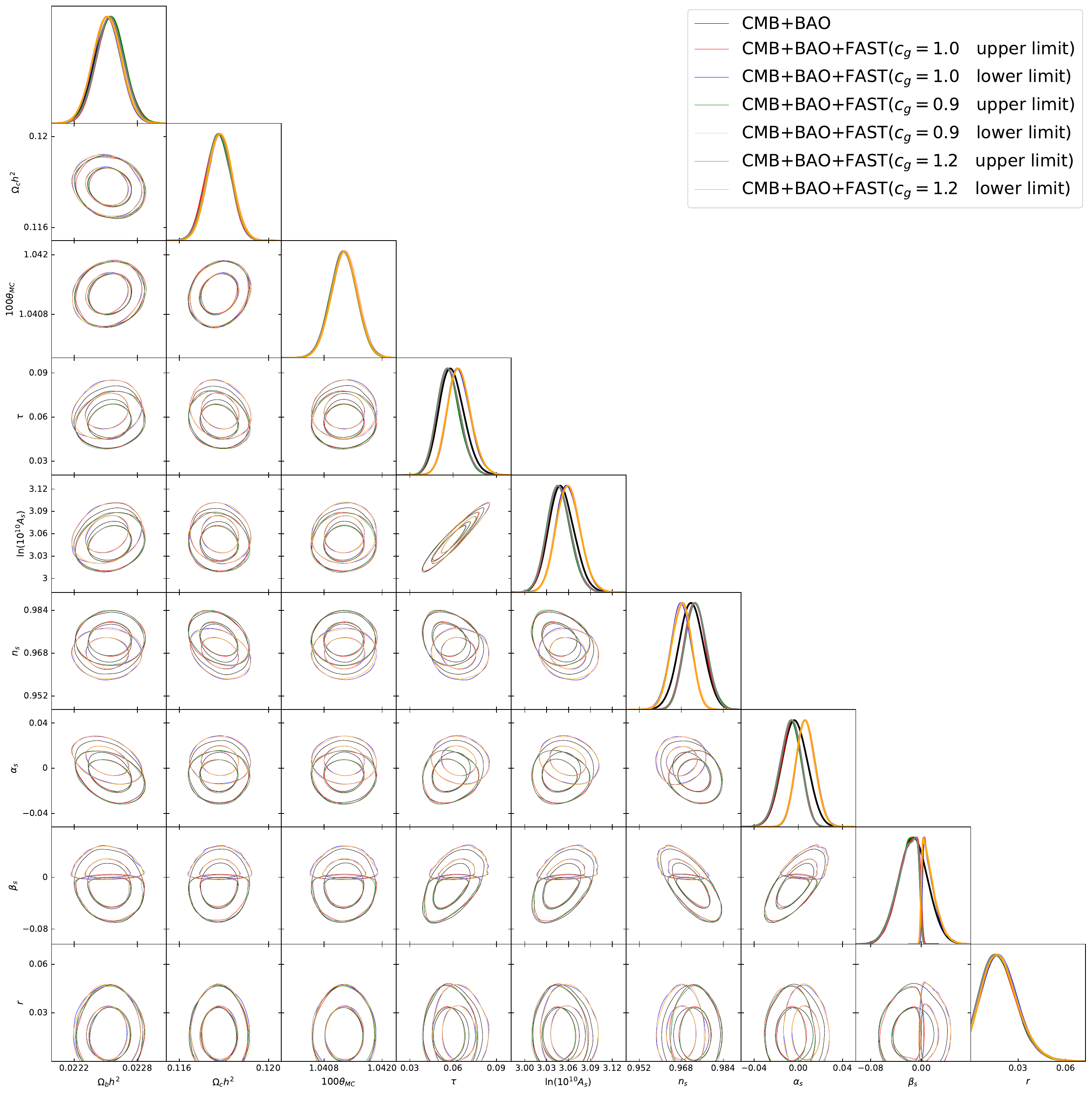}
\caption{The contour plots and likelihood distributions of cosmological parameters in the $\Lambda$CDM+$\alpha_s$+$\beta_s$+$r$ model are shown at the $68\%$ and $95\%$ confidence levels, derived from the combinations of CMB+BAO, CMB+BAO+FAST($c_g=1.0\,$upper limit), CMB+BAO+FAST($c_g=1.0\,$lower limit), CMB+BAO+FAST($c_g=0.9\,$upper limit), CMB+BAO+FAST($c_g=0.9\,$lower limit), CMB+BAO+FAST($c_g=1.2\,$upper limit), and CMB+BAO+FAST($c_g=1.2\,$lower limit) datasets, respectively. The filled lines in the likelihood distributions represent the constraints from CMB+BAO+FAST datasets. The dashed lines in the likelihood distributions represent the constraints from CMB+BAO datasets.}
\label{figure3}
\end{figure}

In the standard $\Lambda$CDM model, the six parameters are the baryon density parameter $\Omega_b h^2$, the cold dark matter density $\Omega_c h^2$, the angular size of the horizon at the last scattering surface $\theta_\text{MC}$, the optical depth $\tau$, the scalar amplitude $A_s$ and the scalar spectral index $n_s$. In the literature, the tensor-to-scalar ratio $r$ is utilized to quantify the tensor amplitude $A_t$ relative to the scalar amplitude $A_s$ at the pivot scale, namely
\e
r\equiv\frac{A_t}{A_s}.
\q
We use the publicly available Cosmomc code \cite{Lewis:2002ah} to extend the standard $\Lambda$CDM model by incorporating parameters such as the tensor-to-scalar ratio $r$, the running of the scalar spectral index $\alpha_s$, and the running of the running of the scalar spectral index $\beta_s$. In the $\Lambda$CDM+r model, the priors for the seven cosmological parameters are set following the default CosmoMC configuration, while the remaining parameters are assigned the following uniform priors: $\alpha_s\in[-0.5,0.5]$ and $\beta_s\in[-0.5, 0.5]$. These parameters are constrained using the combinations of CMB+BAO, along with the upper or lower limits of the stochastic gravitational wave background provided by FAST or SKA. The likelihood for the combined CMB+BAO dataset includes Planck TTTEEE+lowE+lensing, BICEP/Keck 2018 (BK18), 6dF Galaxy Survey, MGS, SDSS DR12, and DESI Data Release 2 (DESI DR2). The upper or lower bounds for the stochastic gravitational-wave background from FAST or SKA are implemented in the subroutine SetFast within the file CosmologyParameterizations.f90 in the source directory. The upper limits on the primordial scalar amplitude $A_s$ inferred from SKA observations at $3.17\times10^{-10}$ Hz are sensitive to the assumed propagation speed of gravitational waves $c_g$. In the $\Lambda$CDM+r framework, the limits read 
\begin{align}
c_g&=1.0:\quad A_s<11.5\times10^{-10}\times(4.09\times10^6)^{1-n_s}, \\
c_g&=0.8:\quad A_s<9.8\times10^{-10}\times(4.09\times10^6)^{1-n_s},  \\
c_g&=1.5:\quad A_s<18.3\times10^{-10}\times(4.09\times10^6)^{1-n_s}.
\end{align}
When the running of the scalar spectral index $\alpha_s$ is included ($\Lambda$CDM+$\alpha_s$+r), the corresponding expressions become
\begin{align}
c_g&=1.0:\quad A_s<11.64\times10^{-10}\times(4.09\times10^6)^{1-n_s-7.61\alpha_s}, \\
c_g&=0.8:\quad A_s<9.84\times10^{-10}\times(4.09\times10^6)^{1-n_s-7.61\alpha_s},  \\
c_g&=1.5:\quad A_s<18.89\times10^{-10}\times(4.09\times10^6)^{1-n_s-7.61\alpha_s}.
\end{align}
Finally, in the extended $\Lambda$CDM+$\alpha_s$+$\beta_s$+r model, the limits derived from FAST data at $6.34\times10^{-10}$ Hz are
\begin{align}
c_g&=1.0:\quad A_s<23.77\times10^{-10}\times(8.18\times10^6)^{1-n_s-7.96\alpha_s-42.23\beta_s}, \\
c_g&=0.9:\quad A_s<10.61\times10^{-10}\times(8.18\times10^6)^{1-n_s-7.96\alpha_s-42.23\beta_s},  \\
c_g&=1.2:\quad A_s<12.66\times10^{-10}\times(8.18\times10^6)^{1-n_s-7.96\alpha_s-42.23\beta_s}.
\end{align}
In the analysis, SKA data are used to constrain the $\Lambda$CDM+r and $\Lambda$CDM+$\alpha_s$+r models; for these parameter spaces, the inclusion of FAST data does not yield stronger limits than the CMB+BAO dataset alone. For the extended $\Lambda$CDM+$\alpha_s$+$\beta_s$+r model, FAST data are employed, as the corresponding SKA limits are much stronger than those obtained from CMB+BAO data. The numerical results are presented in Tables \ref{table1} to \ref{table6}, and Figures \ref{figure1} to \ref{figure3}.

\begin{table*}[thb]
\newcommand{\tabincell}[2]{\begin{tabular}{@{}#1@{}}#2\end{tabular}}
  \centering
  \begin{tabular}{  c |c| c| c}
  \hline
  \hline
  Parameter & \tabincell{c} {CMB+BAO} & \tabincell{c}{CMB+BAO+SKA\\($c_g=1.0\,$upper limit)} & \tabincell{c}{CMB+BAO+SKA\\($c_g=1.0\,$lower limit)}\\
  \hline
  $\Omega_bh^2$ & $0.02253\pm0.00012$   &$0.02247\pm0.00012$ & $0.02253\pm0.00012$ \\
  $\Omega_ch^2$ &$0.11783\pm{0.0006}$    &$0.11850\pm{0.0005}$ & $0.11782\pm{0.0006}$\\
  $100\theta_{\mathrm{MC}}$ & $1.04122\pm0.00027$   &$1.04117\pm0.00027$ & $1.04123\pm0.00027$\\
  $\tau$ &  $0.0619^{+0.0068}_{-0.0079}$   &$0.0527\pm0.0060$  & $0.0616^{+0.0069}_{-0.0081}$\\
  $\ln\(10^{10}A_s\)$  & $3.056^{+0.014}_{-0.016}$     &$3.039^{+0.013}_{-0.012}$ & $3.055^{+0.014}_{-0.016}$\\
  $n_s$ & $0.9698^{+0.0033}_{-0.0032}$  &$0.9598^{+0.0013}_{-0.0009}$ & $0.9697\pm{0.0033}$\\
  $r_{0.05}$  ($95\%$ CL) &$<0.039$  &$<0.038$ & $<0.040$\\
  \hline
  \end{tabular}
  \caption{The $68\%$ confidence limits on the cosmological parameters in the $\Lambda$CDM+$r$ model are derived from the combinations of CMB+BAO, CMB+BAO+SKA($c_g=1.0\,$upper limit), and CMB+BAO+SKA($c_g=1.0\,$lower limit) datasets, respectively.}
  \label{table1}
\end{table*}

\begin{table*}[htb]
\newcommand{\tabincell}[2]{\begin{tabular}{@{}#1@{}}#2\end{tabular}}
  \centering
  \begin{tabular}{  c |c| c|c| c}
  \hline
  \hline
  Parameter & \tabincell{c} {CMB+BAO+SKA\\($c_g=0.8\,$upper limit)} & \tabincell{c}{CMB+BAO+SKA\\($c_g=0.8\,$lower limit)}& \tabincell{c} {CMB+BAO+SKA\\($c_g=1.5\,$upper limit)} & \tabincell{c}{CMB+BAO+SKA\\($c_g=1.5\,$lower limit)}\\
  \hline
  $\Omega_bh^2$ &  $0.02242\pm0.00012$& $0.02253\pm0.00012$ &  $0.02253\pm0.00012$& $0.02265\pm0.00013$\\
  $\Omega_ch^2$ &$0.11912\pm0.0005$&$0.11782\pm{0.0006}$&  $0.11782\pm{0.0006}$& $0.11644\pm{0.0005}$\\
  $100\theta_{\mathrm{MC}}$ &  $1.04112\pm0.00028$& $1.04122\pm0.00028$&  $1.04123\pm0.00027$& $1.04131\pm0.00027$\\
  $\tau$ &    $0.0451^{+0.0070}_{-0.0061}$&  $0.0621^{+0.0070}_{-0.0080}$&  $0.0618^{+0.0069}_{-0.0079}$& $0.0857^{+0.0076}_{-0.0077}$\\
  $\ln\(10^{10}A_s\)$  &  $3.026\pm0.013$& $3.056^{+0.014}_{-0.016}$&  $3.056^{+0.014}_{-0.016}$& $3.099\pm0.015$\\
  $n_s$ &  $0.9506\pm{0.0010}$& $0.9697\pm{0.0033}$&  $0.9698\pm0.0033$& $0.9880\pm0.0011$\\
  $r_{0.05}$  ($95\%$ CL) &$<0.036$&$<0.039$&  $<0.039$& $<0.040$\\
  \hline
  \end{tabular}
  \caption{The $68\%$ confidence limits on the cosmological parameters in the $\Lambda$CDM+$r$ model are derived from the combinations of CMB+BAO+SKA($c_g=0.8\,$upper limit), CMB+BAO+SKA($c_g=0.8\,$lower limit), CMB+BAO+SKA($c_g=1.5\,$upper limit), and CMB+BAO+SKA($c_g=1.5\,$lower limit) datasets, respectively.}
  \label{table2}
\end{table*}

In the $\Lambda$CDM+$r$ model, the scalar spectral index is constrained to $n_s=0.9598^{+0.0013}_{-0.0009}$ from the combinations of CMB+BAO+SKA datasets in the upper limit scenario where scalar-induced gravitational waves propagate at the speed of light. The constraint shifts to $n_s =0.9697\pm{0.0033}$ in the lower limit scenario. When compared to the constraint from the combinations of CMB+BAO datasets alone, the scalar spectral index $n_s$ in the upper limit scenario shows notable changes, suggesting its potential role as an indicator for detecting scalar-induced gravitational waves. Detailed values are provided in Table~\ref{table1}. The contour plots and likelihood distributions of cosmological parameters from the combinations of CMB+BAO+SKA($c_g=1.0\,$lower limit) datasets remain largely consistency with those derived from CMB+BAO. While the contour plots and likelihood distributions from the combinations of CMB+BAO+SKA($c_g=1.0\,$upper limit) datasets noticeably differ from those of CMB+BAO, as depicted in Figure \ref{figure1}. Additionally, we consider scalar-induced gravitational waves propagating at speeds different from the speed of light, as shown in Table~\ref{table2} and Figure~\ref{figure1}. The numerical findings clearly demonstrate the significant influence of the upper and lower limits provided by SKA.

\begin{table*}[htb]
\newcommand{\tabincell}[2]{\begin{tabular}{@{}#1@{}}#2\end{tabular}}
  \centering
  \begin{tabular}{  c |c| c|c}
  \hline
  \hline
  Parameter & \tabincell{c} {CMB+BAO}  & \tabincell{c}{CMB+BAO+SKA\\($c_g=1.0\,$upper limit)}&   \tabincell{c}{CMB+BAO+SKA\\($c_g=1.0\,$lower limit)}\\
  \hline
  $\Omega_bh^2$ & $0.02253\pm0.00014$   &$0.02258\pm0.00013$ & $0.02249\pm0.00013$\\
  $\Omega_ch^2$ &$0.11782\pm0.0006$    &$0.11784\pm0.0006$ & $0.11778\pm0.0006$\\
  $100\theta_{\mathrm{MC}}$ & $1.04123\pm0.00027$      &$1.04123\pm0.00027$ & $1.04121^{+0.00027}_{-0.00028}$\\
  $\tau$ &  $0.0616^{+0.0071}_{-0.0083}$  &$0.0635^{+0.0073}_{-0.0084}$ & $0.0605^{+0.0071}_{-0.0081}$\\
  $\ln\(10^{10}A_s\)$  & $3.055_{-0.017}^{+0.015}$     &$3.060_{-0.016}^{+0.015}$ & $3.052\pm0.015$\\
  $n_s$ & $0.9700\pm{0.0034}$   &$0.9687\pm{0.0034}$& $0.9706^{+0.0034}_{-0.0033}$\\
  $\alpha_s$ &$0.0013\pm0.0080$ &$-0.0071_{-0.0024}^{+0.0060}$& $0.0063^{+0.0032}_{-0.0071}$\\
  $r_{0.05}$  ($95\%$ CL) &$<0.039$   &$<0.039$ & $<0.039$\\
  \hline
  \end{tabular}
  \caption{The $68\%$ confidence limits on the cosmological parameters in the $\Lambda$CDM+$\alpha_s$+$r$ model are derived from the combinations of CMB+BAO, CMB+BAO+SKA($c_g=1.0\,$upper limit), and CMB+BAO+SKA($c_g=1.0\,$lower limit) datasets, respectively.}
  \label{table3}
\end{table*}

\begin{table*}[htb]
\newcommand{\tabincell}[2]{\begin{tabular}{@{}#1@{}}#2\end{tabular}}
  \centering
  \begin{tabular}{  c |c| c|c| c}
  \hline
  \hline
  Parameter & \tabincell{c} {CMB+BAO+SKA\\($c_g=0.8\,$upper limit)} & \tabincell{c}{CMB+BAO+SKA\\($c_g=0.8\,$lower limit)}& \tabincell{c} {CMB+BAO+SKA\\($c_g=1.5\,$upper limit)} & \tabincell{c}{CMB+BAO+SKA\\($c_g=1.5\,$lower limit)}\\
  \hline
  $\Omega_bh^2$ &  $0.02259\pm0.00013$& $0.02250\pm0.00013$ &  $0.02256\pm0.00013$& $0.02248\pm0.00013$\\
  $\Omega_ch^2$ &$0.11784\pm0.0006$&$0.11778\pm0.0006$&  $0.11784\pm0.0006$& $0.11777\pm0.0006$\\
  $100\theta_{\mathrm{MC}}$ &  $1.04122\pm0.00027$& $1.04123^{+0.00027}_{-0.00028}$&  $1.04122\pm0.00027$& $1.04121\pm0.00027$\\
  $\tau$ &    $0.0634^{+0.0073}_{-0.0081}$&  $0.0606^{+0.0070}_{-0.0080}$&  $0.0626^{+0.0070}_{-0.0082}$& $0.0602^{+0.0071}_{-0.0077}$\\
  $\ln\(10^{10}A_s\)$  &  $3.061\pm0.015$& $3.052\pm0.015$&  $3.058^{+0.014}_{-0.017}$& $3.051\pm0.015$\\
  $n_s$ &  $0.9687\pm0.0033$& $0.9707\pm0.0034$&  $0.9691\pm0.0033$& $0.9711\pm0.0033$\\
  $\alpha_s$ &$-0.0078_{-0.0021}^{+0.0053}$ &$0.0058_{-0.0078}^{+0.0036}$& $-0.0041^{+0.0069}_{-0.0030}$&  $0.0088_{-0.0059}^{+0.0024}$\\
  $r_{0.05}$  ($95\%$ CL) &$<0.039$&$<0.037$&  $<0.039$& $<0.039$\\
  \hline
  \end{tabular}
  \caption{The $68\%$ confidence limits on the cosmological parameters in the $\Lambda$CDM+$\alpha_s$+$r$ model are derived from the combinations of CMB+BAO+SKA($c_g=0.8\,$upper limit), CMB+BAO+SKA($c_g=0.8\,$lower limit), CMB+BAO+SKA($c_g=1.5\,$upper limit), and CMB+BAO+SKA($c_g=1.5\,$lower limit) datasets, respectively.}
  \label{table4}
\end{table*}

In the $\Lambda$CDM+$\alpha_s$+$r$ model, the scalar spectral index and the running of the scalar spectral index are constrained to $n_s=0.9687\pm{0.0034}$ and $\alpha_s=-0.0071_{-0.0024}^{+0.0060}$, respectively, from the combinations of CMB+BAO+SKA datasets in the upper limit scenario where scalar-induced gravitational waves propagate at the speed of light. The constraints shift to $n_s=0.9706^{+0.0034}_{-0.0033}$ and $\alpha_s=0.0063^{+0.0032}_{-0.0071}$ in the lower limit scenario. Comparison with constraints from the combinations of CMB+BAO datasets reveals notable changes in $\alpha_s$. Detailed values can be found in Table~\ref{table3}. The contour plots and likelihood distributions of $\alpha_s$ indicate a predominantly positive constraint in the lower limit scenario, whereas the constraints lean negative in the upper limit scenario, as depicted in Figure \ref{figure2}. Additionally, we consider scalar-induced gravitational waves propagating at speeds different from the speed of light, as shown in Table~\ref{table4} and Figure \ref{figure2}.

\begin{table*}[htb]
\newcommand{\tabincell}[2]{\begin{tabular}{@{}#1@{}}#2\end{tabular}}
  \centering
  \begin{tabular}{  c |c| c|c}
  \hline
  \hline
  Parameter & \tabincell{c} {CMB+BAO} & \tabincell{c}{CMB+BAO+FAST\\($c_g=1.0\,$upper limit)} & \tabincell{c}{CMB+BAO+FAST\\($c_g=1.0\,$lower limit)}\\
  \hline
  $\Omega_bh^2$ & $0.02254^{+0.00013}_{-0.00014}$ &$0.02255\pm0.00013$ & $0.02251\pm0.00013$\\
  $\Omega_ch^2$ &$0.11779\pm{0.0006}$  &$0.11775\pm{0.0006}$ & $0.11783\pm0.0006$\\
  $100\theta_{\mathrm{MC}}$ & $1.04121\pm0.00027$    &$1.04122\pm0.00027$ & $1.04122^{+0.00028}_{-0.00027}$\\
  $\tau$ &  $0.0591^{+0.0081}_{-0.0095}$  &$0.0571^{+0.0072}_{-0.0088}$ & $0.0642^{+0.0076}_{-0.0087}$\\
  $\ln\(10^{10}A_s\)$  & $3.051^{+0.016}_{-0.019}$      &$3.047^{+0.015}_{-0.018}$ & $3.059^{+0.016}_{-0.018}$\\
  $n_s$ & $0.9718\pm{0.0047}$    &$0.9733\pm0.0041$ & $0.9679\pm{0.0039}$\\
  $\alpha_s$ &$-0.0033^{+0.012}_{-0.011}$ &$-0.0068_{-0.009}^{+0.010}$ & $0.0068\pm{0.009}$\\
  $\beta_s$ &$-0.0131\pm0.023$ &$-0.0232^{+0.021}_{-0.010}$ & $0.0147^{+0.006}_{-0.015}$\\
  $r_{0.05}$  ($95\%$ CL) &$<0.039$   &$<0.038$ & $<0.039$\\
  \hline
  \end{tabular}
  \caption{The $68\%$ confidence limits on the cosmological parameters in the $\Lambda$CDM+$\alpha_s$+$\beta_s$+$r$ model are derived from the combinations of CMB+BAO, CMB+BAO+FAST($c_g=1.0\,$upper limit), and CMB+BAO+FAST($c_g=1.0\,$lower limit) datasets, respectively.}
  \label{table5}
\end{table*}

\begin{table*}[htb]
\newcommand{\tabincell}[2]{\begin{tabular}{@{}#1@{}}#2\end{tabular}}
  \centering
  \begin{tabular}{  c |c| c|c| c}
  \hline
  \hline
  Parameter & \tabincell{c} {CMB+BAO+FAST\\($c_g=0.9\,$upper limit)} & \tabincell{c}{CMB+BAO+FAST\\($c_g=0.9\,$lower limit)}& \tabincell{c} {CMB+BAO+FAST\\($c_g=1.2\,$upper limit)} & \tabincell{c}{CMB+BAO+FAST\\($c_g=1.2\,$lower limit)}\\
  \hline
  $\Omega_bh^2$ &  $0.02254^{+0.00013}_{-0.00014}$& $0.02251\pm0.00014$ &  $0.02254\pm0.00013$& $0.02251\pm0.00013$\\
  $\Omega_ch^2$ &$0.11777\pm0.0006$&$0.11783\pm{0.0006}$&  $0.11777\pm0.0006$& $0.11782^{+0.0005}_{-0.0006}$\\
  $100\theta_{\mathrm{MC}}$ &  $1.04120\pm0.00027$& $1.04122^{+0.00027}_{-0.00028}$&  $1.04121\pm0.00027$& $1.04122\pm0.00027$\\
  $\tau$ &    $0.0570_{-0.0088}^{+0.0072}$&  $0.0642_{-0.0089}^{+0.0078}$&  $0.0571^{+0.0079}_{-0.0087}$& $0.0643^{+0.0075}_{-0.0089}$\\
  $\ln\(10^{10}A_s\)$  &  $3.047_{-0.017}^{+0.015}$& $3.060_{-0.017}^{+0.016}$&  $3.047_{-0.017}^{+0.016}$& $3.060_{-0.018}^{+0.016}$\\
  $n_s$ &  $0.9735\pm0.0042$& $0.9680\pm0.0039$&  $0.9734\pm0.0042$& $0.9680_{-0.0038}^{+0.0041}$\\
  $\alpha_s$ &$-0.0073_{-0.009}^{+0.011}$ &$0.0065_{-0.009}^{+0.008}$ & $-0.0072^{+0.011}_{-0.009}$&$0.0065\pm0.009$ \\
  $\beta_s$ &$-0.0243_{-0.010}^{+0.022}$ &$0.0144_{-0.015}^{+0.006}$ & $-0.0242^{+0.022}_{-0.010}$&$0.0142_{-0.015}^{+0.006}$ \\
  $r_{0.05}$  ($95\%$ CL) &$<0.038$&$<0.039$&  $<0.038$& $<0.039$\\
  \hline
  \end{tabular}
  \caption{The $68\%$ confidence limits on the cosmological parameters in the $\Lambda$CDM+$\alpha_s$+$\beta_s$+$r$ model are derived from the combinations of CMB+BAO+FAST($c_g=0.9\,$upper limit), CMB+BAO+FAST($c_g=0.9\,$lower limit), CMB+BAO+FAST($c_g=1.2\,$upper limit), and CMB+BAO+FAST($c_g=1.2\,$lower limit) datasets, respectively.}
  \label{table6}
\end{table*}

In the $\Lambda$CDM+$\alpha_s$+$\beta_s$+$r$ model, the parameters describing the scalar spectral index, the running of the scalar spectral index, and the running of the running of the scalar spectral index are constrained as follows: $n_s=0.9733\pm0.0041$, $\alpha_s=-0.0068_{-0.009}^{+0.010}$, and $\beta_s=-0.0232^{+0.021}_{-0.010}$, derived from the combinations of CMB+BAO+FAST datasets in the upper limit scenario where scalar-induced gravitational waves propagate at the speed of light. The values shift to $n_s=0.9679\pm{0.0039}$, $\alpha_s=0.0068\pm{0.009}$ ,and $\beta_s=0.0147^{+0.006}_{-0.015}$ in the lower limit scenario. Comparative analysis with constraints solely from CMB+BAO datasets demonstrates significant variations, particularly in $\alpha_s$ and $\beta_s$. Detailed numerical values are summarized in Table~\ref{table5}. The contour plots and likelihood distributions of $\beta_s$ illustrate predominantly positive constraints in the lower limit scenario, whereas they lean negative in the upper limit scenario, as depicted in Figure \ref{figure3}. Additionally, we consider scalar-induced gravitational waves propagating at speeds different from the speed of light, as shown in Table~\ref{table6} and Figure \ref{figure3}. The findings underscore the significant impact of upper and lower limits provided by FAST. In the $\Lambda$CDM+$\alpha_s$+$r$ model and the $\Lambda$CDM+$\alpha_s$+$\beta_s$+$r$ model, the running of the scalar spectral index $\alpha_s$ and the running of the running of the scalar spectral index $\beta_s$ exhibit notable variations, indicating potential observational insights.

\section{summary}
Scalar-induced gravitational waves originating from the early universe are a key prediction of various inflationary models where quantum fluctuations during inflation can generate a stochastic background of gravitational waves. These waves imprint specific signatures in the polarization of CMB and in the large-scale structure of the universe. The signatures offer complementary observational avenues to confirm their existence and investigate their properties. Detecting scalar-induced gravitational waves requires highly sensitive instruments capable of measuring minute distortions in spacetime.

In this paper, we combined CMB+BAO datasets with upper or lower limits of the stochastic gravitational wave background provided by FAST or SKA, to constrain scalar-induced gravitational waves. In the $\Lambda$CDM+$r$ model, the scalar spectral index of the power-law power spectrum is constrained to $n_s=0.9598^{+0.0013}_{-0.0009}$ in the upper limit scenario, derived from combined CMB+BAO+SKA datasets where scalar-induced gravitational waves propagate at the speed of light. In the lower limit scenario, this constraint shifts to $n_s = 0.9697\pm{0.0033}$. These constraints in the upper limit scenario represent significant deviations from those obtained using CMB+BAO datasets alone, suggesting $n_s$ as a potential indicator for detecting scalar-induced gravitational waves. In the $\Lambda$CDM+$\alpha_s$+$r$ model and the $\Lambda$CDM+$\alpha_s$+$\beta_s$+$r$ model, the running of the scalar spectral index $\alpha_s$ and the running of the running of the scalar spectral index $\beta_s$ also exhibit notable variations. These parameters further underscore the sensitivity of cosmological constraints to gravitational wave signatures, particularly influenced by the upper and lower limits provided by facilities like FAST or SKA. These numerical results emphasize the critical role of future observatories in refining our understanding of early universe physics through precise measurements of $n_s$, $\alpha_s$, and $\beta_s$ from combined datasets. Such advancements promise to deepen our insights into the origin and evolution of the cosmos.

\noindent {\bf Acknowledgments}.
This work is supported by the National Natural Science Foundation of China (Grant No. 12405069), the Natural Science Foundation of Shandong Province (Grant No. ZR2021QA073) and Research Start-up Fund of QUST (Grant No. 1203043003587).


\end{document}